\documentclass[aps,pra,floatfix,twocolumn,showpacs]{revtex4}
\usepackage{epsfig}
\usepackage{graphicx}
\usepackage{dcolumn}
\usepackage{bm}
\begin{document}
\title{\bf Breit Interaction and Parity Non-conservation in Many-Electron Atoms}
\author{V. A. Dzuba}
\email{V.Dzuba@unsw.edu.au}
\author{V. V. Flambaum}
\email{V.Flambaum@unsw.edu.au}
\affiliation{School of Physics, University of New South Wales, 
Sydney 2052, Australia}
\author{M. S. Safronova}
\email{msafrono@udel.edu}
\affiliation{Department of Physics and Astronomy,
University of Delaware, Newark, Delaware 19716, USA}
\date{\today}

\begin{abstract}

We present accurate {\em ab initio} non-perturbative calculations of the 
Breit correction to the parity non-conserving (PNC) amplitudes of 
the $6s-7s$ and $6s-5d_{3/2}$ transitions in Cs, $7s-8s$ and 
$7s-6d_{3/2}$ transitions in Fr,
$6s-5d_{3/2}$ transition in Ba$^+$, 
$7s-6d_{3/2}$ transition in Ra$^+$, 
and $6p_{1/2} - 6p_{3/2}$ transition in Tl.
The results for the $6s-7s$ transition in Cs and $7s-8s$  
transition in Fr are in good agreement with other calculations
while calculations for other atoms/transitions are presented 
for the first time.
We demonstrate that higher-orders many-body corrections to
the Breit interaction are especially important
for the $s-d$ PNC amplitudes.
We confirm good agreement of the PNC
measurements for cesium and thallium with the standard model .

\end{abstract} 

\pacs{PACS: 32.80.Ys,31.30.Jv,11.30.Er,12.15.Ji}

\maketitle

\section{Introduction}

Study of the parity non-conservation (PNC) in atoms have reached the accuracy
where small corrections like Breit interaction and radiative corrections
play significant role (see, e.q. review \cite{Ginges}). For example,
experimental accuracy for the PNC in cesium is 0.35\% \cite{Wood}.
Interpretation of this result using earlier atomic calculations 
\cite{Dzuba89,Blundell90} lead to apparent disagreement with the standard
model~\cite{Bennett}. This disagreement had been resolved when 
Breit~\cite{Derevianko,Harabati,Kozlov} and 
radiative~\cite{Flambaum,Sushkov,Shabaev} corrections were included
into analysis.

In present paper we revisit Breit corrections to the PNC amplitudes in
many-electron atoms. Previous calculations were focused on the  
$6s-7s$ PNC amplitude in Cs~\cite{Derevianko,Harabati,Kozlov,Shabaev1}, 
while some also considered very similar (in terms of electron structure)
$7s-8s$ PNC amplitude in Fr~\cite{Derevianko3,Safronova,Shabaev}.
The results are considered to be in reasonable agreement with each
other in spite of some difference in the final values given by
different calculations (up to $\sim 30\%$). This is because
Breit correction to the PNC amplitude in cesium was small ($\sim -0.6\%$)
and relatively rough estimations were sufficient for the sake
of the interpretation of the experimental measurements.

However, if we need to calculate Breit corrections to PNC in atoms
heavier than cesium or, in general, if we want to have a reliable
method to study the effect of Breit interaction in heavy many-electron
atoms the important question to ask is what kind of approach should
be used to do this? The answer is evident from the analysis of the
details of the calculations of Breit contributions by means of
many-body perturbation theory (MBPT). The analysis reveals very 
poor convergence of the MBPT with respect to the number of
residual Coulomb 
interactions included into the higher-order terms. This means that
some all-order technique is needed to sum up important chains of 
higher-order terms to all orders. The natural choice is to
present inter-electron interaction as a sum of Coulomb and Breit
terms and to treat both terms the same way.

We have developed such approach in our previous
work~\cite{Harabati} and applied it to the analysis of the
PNC in cesium. Breit and Coulomb interaction were equally
treated on the self-consistent field level (Hartree-Fock
and random-phase approximation calculations). However,
inclusion of the Breit interaction into calculation of
the correlations was not comprehensive. In that work we
included correlations by calculating the correlation
potential $\hat \Sigma$ for valence electrons, using it
to calculate the so-called Brueckner orbitals for the states
of valence electrons, and then replacing Hartree-Fock
orbitals by Brueckner orbitals in the $E1_{PNC}$ amplitude.
Brueckner orbitals were calculated with Breit interaction
included into self-consistent Hartree-Fock potential while
correlation potential $\hat \Sigma$ was calculated without
inclusion of the Breit interaction.

In present work we do one more step by including Breit interaction
into second-oder correlation potential $\hat \Sigma$ thus
achieving equal and comprehensive treatment of Coulomb and Breit
interactions on levels up to second order of the many-body
perturbation theory.
We apply this approach to the calculation of the Breit corrections 
to the PNC amplitudes in neutral cesium, thallium and francium as well 
as barium and radium positive ions. This is the full list of atoms and
ions with one external electron above closed shells for which PNC
measurements were done or planned. Our calculations represent
the most comprehensive treatment of the Breit contribution
to the PNC in many-electron atoms. The results for Tl, Ba$^+$,
Ra$^+$ as well as for the $s-d$ transitions for Cs and Fr are 
presented for the first time.

The method developed in present work is important not only for
the PNC in many-electron atoms. It can be equally useful for calculation
of Breit interaction contribution to any observable values enhanced 
on short distances. These include magnetic dipole and electric quadrupole
hyperfine structure, isotope shift, T-odd effects, etc.

\section{Theory}

\begin{figure*}
\centering
\epsfig{figure=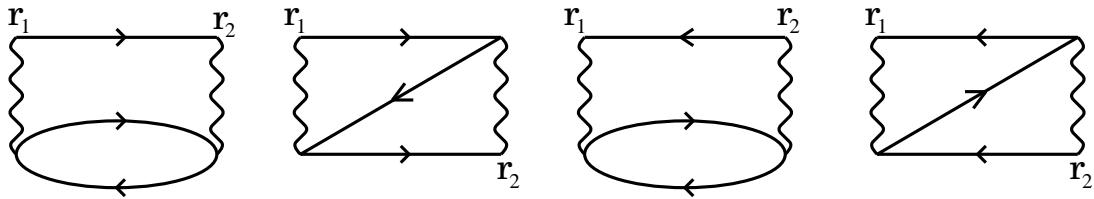,scale=0.7}
\caption{Second-order correlation potential $\hat \Sigma(r_1,r_2)$.
Left arrows denote core states, right arrows denote states above the core,
waved line is the sum of Coulomb and Breit interactions.}
\label{Sigma2}
\end{figure*}

We use the following form of the Breit operator \cite{Breit}
(atomic units are used throughout the paper)
\begin{equation}
	\hat H^B = - \frac{{\bf \alpha_1}\cdot{\bf \alpha_2}+
	({\bf \alpha_1}\cdot{\bf \hat{n}})({\bf \alpha_2}
	\cdot{\bf \hat{n}})}{2r}.
\label{HBreit}
\end{equation}
Here ${\bf r} = {\bf \hat{n}}r$, $r$ is distance between electrons and 
$\alpha_i$ is $\alpha$-matrix of the corresponding electron. This is a
low frequency limit of the relativistic correction to the Coulomb interaction 
between electrons. It contains magnetic interaction and 
retardation.

Similar to Coulomb interaction, Breit interaction creates self-consistent
Breit potential
\begin{eqnarray}
	\hat V^B \psi(r) = \sum_n (
 \int{\psi_n^{\dagger} \hat H^B \psi_n d^3 r'} \, \psi(r)  - \nonumber \\
	\int{\psi_n ^{\dagger}\hat H^B \psi d^3 r'} \, \psi_n(r)).
\label{VBreit}
\end{eqnarray}
Here summation goes over all core states. In the case of closed-shell atoms
(like alkali atoms in the $V^{N-1}$ approximation) direct term in Breit
potential vanishes (first term in~(\ref{VBreit})) and only second, exchange,
term remains.

Hartree-Fock Hamiltonian has the form
\begin{equation}
	\hat H_0 =  c\bm{\alpha p} + (\beta-1)mc^2
     - \frac{Ze^2}{r} + \hat V,
\label{HFG}
\end{equation}
where $\hat V$ is the self-consistent potential created by electrons 
from the core. To achieve comprehensive treatment of the Breit
interaction potential $\hat V$ should be presented as a sum of
Coulomb and Breit terms
\begin{equation}
  \hat V = \hat V^C + \hat V^B.
\label{VCB}
\end{equation}
Note that we write for simplicity nuclear potential in (\ref{HFG})
as for point-like nucleus. However, our actual potential takes into account
finite nuclear size.

To do PNC calculations one needs to take into account dipole
interaction of an atom with external photon as well as weak interaction
of atomic electrons with the nucleus. We use the time-dependent Hartree-Fock
method (TDHF) to do this~\cite{CPM}. 
This method is equivalent to well-known random-phase approximation (RPA).

For every electron in the atom, single electron wave function is presented
in the form
\begin{eqnarray}
  \tilde \psi_n &=& \psi_n + \delta \psi_n + X_a e^{-i\omega t} + Y_n e^{i\omega t}
\nonumber \\
 &+& \delta X_n e^{-i\omega t} + \delta Y_n e^{i\omega t},
 \label{psi}
\end{eqnarray}
where index $n$ numerates single-electron states, $\psi_n$ is unperturbed
wave function for the state $n$ which is an eigenstate of the Hartree-Fock
Hamiltonian~(\ref{HFG}), $\delta \psi_n$ is correction to $\psi_n$ due
to weak interaction with the nucleus, $X_n$ and $Y_n$ are corrections due to
electric field of external photon of frequency $\omega$, $\delta X_n$ and
$\delta Y_n$ are corrections due to simultaneous action of the weak interaction
and dipole interaction with external photon.

Corrections $\delta \psi_n$, $X_n$, $Y_n$, $\delta X_n$, $\delta Y_n$ 
to all atomic states found by self-consistent iteration of the TDHF
equations:
\begin{eqnarray}
  (\hat H_0 - \epsilon_n)\delta \psi_n &=& -(\hat H_W + \delta \hat V_W)\psi_n,\label{weak} \\
  (\hat H_0 - \epsilon_n - \omega)X_n &=& -(\hat H_{E1} + \delta \hat V_{E1})\psi_n,\nonumber \\
  (\hat H_0 - \epsilon_n + \omega)Y_n &=& -(\hat H_{E1}^{\dagger} + 
  \delta \hat V_{E1}^{\dagger})\psi_n, \label{E1} \\
  (\hat H_0 - \epsilon_n - \omega)\delta X_n &=& 
 - \delta \hat V_{E1}\delta \psi_n \nonumber \\
 &-& \delta \hat V_{W}X_n 
 - \delta \hat V_{E1W}\psi_n, \nonumber \\
  (\hat H_0 - \epsilon_n + \omega)\delta Y_n &=& 
 - \delta \hat V_{E1}^{\dagger}\delta \psi_n \label{E1W} \\
 &-& \delta \hat V_{W}Y_n 
 - \delta \hat V_{E1W}^{\dagger}\psi_n. \nonumber
\end{eqnarray}
Here $\delta \hat V$ is correction to the self-consistent Hartree-Fock
potential $\hat V$ due to external field (weak ($W$), dipole electric ($E1$)
or both ($E1W$)). Equations (\ref{weak},\ref{E1},\ref{E1W}) are first 
solved self-consistently for states in the core. 
Then corrections to valence states are calculated
in the field of frozen core.

PNC amplitude for the transition between valence states $a$ and $b$ in
the TDHF approximation is
\begin{eqnarray}
  E_{PNC}^{TDHF}& = &
  \langle \psi_b | \hat H_{E1} + \delta \hat V_{E1}|\delta \psi_a \rangle
\label{TDHF} \\ & + &
  \langle \psi_b | \hat H_{W}  + \delta \hat V_{W} | X_a \rangle +
  \langle \psi_b | \delta \hat V_{E1W}| \psi_a \rangle. \nonumber
\end{eqnarray}
Breit interaction is included by using expression (\ref{VCB}) for the self-consistent
potential on all stages of the calculations. In other words, Hartree-Fock
potential in the left-hand side of equations (\ref{weak},\ref{E1},\ref{E1W}) is 
the sum of Coulomb and Breit terms. Correction to the potential $\delta \hat V$
in the right-hand side of equations (\ref{weak},\ref{E1},\ref{E1W},\ref{TDHF}) is
also the sum of Coulomb and Breit terms:
\[  \delta \hat V =  \delta \hat V^C + \delta \hat V^B. \]
Expression (\ref{TDHF})  does not take into account correlations.
To include correlations we introduce correlation potential
$\hat \Sigma$. It is defined in such a way that its average value
over particular valence state $a$ is the correlation correction
to the energy of this state (see, e.g.~\cite{CPM} for details):
\begin{equation}
  \Delta \epsilon_a = \langle a | \hat \Sigma | a \rangle. 
\label{Sigma}
\end{equation}
$\hat \Sigma$ is a non-local operator which depends on energy
and parameter $\kappa$ which defines the angular part of wave
function $\psi_a$: $\kappa_a = (-1)^{l_a+j_a+1/2}(j_a+1/2)$,
where $l_a$ is angular momentum and $j_a$ is total momentum
of state $a$.

In linear in $\hat \Sigma$ approximation, correlation correction
to the PNC amplitude is
\begin{eqnarray}
 && \Delta E_{PNC} = 
  \langle \psi_b | \hat \Sigma_{\kappa_b}(\epsilon_b)|\delta X_a \rangle +
  \langle \delta \psi_b | \hat \Sigma_{-\kappa_b}(\epsilon_b)|X_a \rangle 
\nonumber \\
&& + \langle \delta Y_b | \hat \Sigma_{\kappa_a}(\epsilon_a)|\psi_a \rangle +
  \langle Y_b | \hat \Sigma_{-\kappa_a}(\epsilon_a)|\delta \psi_a \rangle.
 \label{cor}
\end{eqnarray}
In the most precise calculations of the PNC in cesium~\cite{Dzuba89,Dzuba02},
operator $\hat \Sigma$ was calculated in all orders in Coulomb interaction.
For the purpose of the present work we can restrict ourself with the
lowest, second-order expression for $\hat \Sigma$. Corresponding
Brueckner-Goldstone diagrams are presented on Fig.~\ref{Sigma2}.
Breit interaction is included in calculation of $\hat \Sigma$ in two
ways. First, a complete set of single-electron states used to
calculate $\hat \Sigma$ was obtained with the use of the 
Hartree-Fock Hamiltonian (\ref{HFG}) with Breit potential
included as in eq.~(\ref{VCB}). Second, inter-electron interaction
in $\hat \Sigma$ (wave lines in all diagrams on Fig.~\ref{Sigma2})
is the sum of Coulomb ($e^2/r$) and Breit (\ref{HBreit}) terms.  

The total PNC amplitude in linear in $\hat \Sigma$ approximation is 
\begin{equation}
  E_{PNC} =   E_{PNC}^{TDHF} +  \Delta E_{PNC}.
 \label{final}
\end{equation}
We also use a more accurate way of calculating correlation corrections
which includes correlation potential $\hat \Sigma$ to all orders.
We do this with the help of so-called Brueckner orbitals~\cite{CPM}.
Brueckner orbitals are found by including $\hat \Sigma$
into Hartree-Fock equations for valence states. 
For example, for the states $a$ and $b$ 
Brueckner orbitals $\psi^{Br}_a$ and $\psi^{Br}_a$ and corresponding energies
$\epsilon_a$ and $\epsilon_b$ are found by solving
\begin{eqnarray}
  (\hat H_0 + \hat \Sigma_{\kappa_a} -\epsilon_a)\psi^{Br}_a &=& 0, 
\nonumber \\
  (\hat H_0 + \hat \Sigma_{\kappa_b} -\epsilon_b)\psi^{Br}_b &=& 0.
\label{Brab}
\end{eqnarray}
Corrections $\delta \psi^{Br}_a$, $X^{Br}_a$ and $Y^{Br}_a$ to
Brueckner orbital $\psi^{Br}_a$ induced by weak and electromagnetic
interactions are found by solving
\begin{eqnarray}
  (\hat H_0 + \hat \Sigma_{-\kappa_a} - \epsilon_a)\delta \psi^{Br}_a = 
 -(\hat H_W + \delta \hat V_W)\psi^{Br}_a, \nonumber \\
  (\hat H_0 + \hat \Sigma_{-\kappa_b} - \epsilon_a - \omega)X^{Br}_a = 
  -(\hat H_{E1} + \delta \hat V_{E1})\psi^{Br}_a,\label{DBrab} \\
  (\hat H_0 + \hat \Sigma_{-\kappa_b}- \epsilon_a + \omega)Y^{Br}_a = 
 -(\hat H_{E1}^{\dagger} + 
  \delta \hat V_{E1}^{\dagger})\psi^{Br}_a. \nonumber 
\end{eqnarray}
Then, the all-order in $\hat \Sigma$ PNC amplitude is given by the expression 
similar to (\ref{TDHF}) but with Hartree-Fock orbitals replaces by Brueckner 
orbitals:
\begin{eqnarray}
  E_{PNC}^{Br}& = &
  \langle \psi^{Br}_b | \hat H_{E1} + \delta \hat V_{E1}|\delta \psi^{Br}_a \rangle
\label{Ecor} \\ & + &
  \langle \psi^{Br}_b | \hat H_{W}  + \delta \hat V_{W} | X^{Br}_a \rangle +
  \langle \psi^{Br}_b | \delta \hat V_{E1W}| \psi^{Br}_a \rangle. \nonumber
\end{eqnarray}
Equations (\ref{final}) and (\ref{Ecor}) differ by non-linear in $\hat \Sigma$
terms ($\hat \Sigma^2$, $\hat \Sigma^3$, etc.).

Finally we note that non-perturbative treatment of 
Breit interaction leads to inclusion of higher-order in Breit operator
terms, terms proportional to $(\hat H^B)^2$,$(\hat H^B)^3$, etc. 
with $\hat H^B$ given by (\ref{HBreit}). 
Inclusion of these terms cannot be justified 
since expression (\ref{HBreit}) is already approximate.
Non-linear terms can be easily eliminated by a rescaling
procedure. In this procedure inter-electron interaction is
considered as a sum $U(r)=e^2/r + \lambda \hat H^B$ in which
Breit interaction $\hat H^B$ is multiplied by a scaling parameter
$\lambda$. By running programs for different values of $\lambda$
a range of values is found for which final answer is linear
function of $\lambda$. Then the answer is interpolated to
$\lambda=1$. It turns out, however, that as a rule $\lambda=1$
is already in linear regime.

\section{Results and discussion}

\begin{table*}
\caption{Contribution of Breit interaction ($\Delta$) into $E_{PNC}$ 
transition amplitudes in Cs, Ba$^+$, Tl, Fr 
and Ra$^+$ ($10^{-11} iea_B Q_W/N$)}
\label{PNC}
\begin{ruledtabular}
\begin{tabular}{lccccccc}
Atom & $Z$ & Transition & Approximation & Coulomb only & 
Coulomb + Breit & $\Delta$ & $\Delta/E_{PNC}$(\%) \\
\hline
~Cs & 55 & $6s_{1/2} - 7s_{1/2}$ & TDHF\footnotemark[1] & 0.8903 & 0.8849 & -0.0054 & -0.61 \\
    &    &        & $\hat \Sigma^{(2)}$\footnotemark[2] & 0.9316 & 0.9258 & -0.0058 & -0.62 \\
    &    &        & Br\footnotemark[3]                  & 0.9001 & 0.8945 & -0.0056 & -0.62 \\
    &    &        &                                     &        &        &         &       \\
    &    & $6s_{1/2} - 5d_{3/2}$ & TDHF\footnotemark[1] & 3.1177 & 3.1012 & -0.0165 & -0.53 \\
    &    &        & $\hat \Sigma^{(2)}$\footnotemark[2] & 3.5371 & 3.5195 & -0.0176 & -0.50 \\
    &    &        & Br\footnotemark[3]                  & 3.7277 & 3.7165 & -0.0112 & -0.30 \\
    &    &        &                                     &        &        &         &       \\
~Ba$^+$
   & 56 & $6s_{1/2} - 5d_{3/2}$ & TDHF\footnotemark[1] & 2.1678  & 2.1513 & -0.0165 & -0.76 \\
   &    &        & $\hat \Sigma^{(2)}$\footnotemark[2] & 2.2091  & 2.1915 & -0.0176 & -0.80 \\
   &    &        & Br\footnotemark[3]                  & 2.1826  & 2.1653 & -0.0173 & -0.79 \\
    &    &        &                                     &        &        &         &       \\
~Tl & 81 & $6p_{1/2} - 6p_{3/2}$ & TDHF\footnotemark[1] & 30.384  & 30.105 & -0.279  & -0.92 \\
   &    &         & $\hat \Sigma^{(2)}$\footnotemark[2] & 23.157  & 22.941 & -0.216  & -0.93 \\
   &    &         & Br\footnotemark[3]                  & 24.270  & 24.046 & -0.224  & -0.92 \\
    &    &        &                                     &        &        &         &       \\
~Fr & 87 & $7s_{1/2} - 8s_{1/2}$ & TDHF\footnotemark[1] & 16.043  & 15.891 & -0.152  & -0.95 \\
    &    &        & $\hat \Sigma^{(2)}$\footnotemark[2] & 15.884  & 15.729 & -0.155  & -0.98 \\
    &    &        & Br\footnotemark[3]                  & 15.229  & 15.079 & -0.150  & -0.98 \\
    &    &        &                                     &        &        &         &       \\
    &    & $7s_{1/2} - 6d_{3/2}$ & TDHF\footnotemark[1] & 52.118  & 51.678 & -0.440  & -0.84 \\
    &    &        & $\hat \Sigma^{(2)}$\footnotemark[2] & 55.568  & 55.103 & -0.465  & -0.84 \\
    &    &        & Br\footnotemark[3]                  & 56.437  & 56.090 & -0.347  & -0.61 \\
    &    &        &                                     &        &        &         &       \\
~Ra$^+$
    & 88 & $7s_{1/2} - 6d_{3/2}$ & TDHF\footnotemark[1] & 44.038  & 43.484  & -0.554 & -1.26 \\
    &    &        & $\hat \Sigma^{(2)}$\footnotemark[2] & 42.817  & 42.269  & -0.548 & -1.29 \\
    &    &        & Br\footnotemark[3]                  & 42.511  & 41.971  & -0.540 & -1.27 \\
\end{tabular}
\footnotetext[1]{TDHF: time-dependent Hartree-Fock, Eq.(\ref{TDHF})}
\footnotetext[2]{$\hat \Sigma^{(2)}$: second-order correlations are included, Eq.(\ref{final})}
\footnotetext[3]{Br: Brueckner orbitals, Eq.~(\ref{Ecor})}
\end{ruledtabular}
\end{table*}

Table~\ref{PNC} presents results of calculations of the Breit
correction to the PNC amplitudes in
Cs, Ba, Tl, Fr and Ra.
This is the full list of atoms with one external electron above 
closed shells for which PNC measurement has been carried out or planned.
Comparison with other calculations for the $6s -7s$ transition in 
Cs and  $7s -8s$ transition in Fr are presented
in Table~\ref{comp}. 
The results for other atoms and for the $s-d$ transitions in Cs and Fr
are presented for the first time.

Note that for all PNC amplitudes presented in Table~\ref{PNC} the
effect of correlations on Breit correction is small in $\hat \Sigma^{(2)}$
approximation. It becomes significant for some s-d amplitudes only
in Brueckner approximation.
This is because the correlation corrections in d-wave are large
and significantly change density of the valence electron
inside electron core. Therefore, we should perform
calculations using  Bruckner orbitals which include  $\hat \Sigma$
to all orders. The linear in $\hat \Sigma$
expressions (\ref{cor}),(\ref{final}) do not provide 
satisfactory accuracy.

As one can see from Table~\ref{PNC} the behavior of $s-s$ and $s-d$
PNC amplitudes is very different. Let us discuss them separately.
It is convenient to use the sum-over-states expression for the E1
PNC amplitude for the discussion:
\begin{eqnarray}
  E1_{PNC}(ab) &=& \sum_n \frac{\langle b|\hat D|n\rangle\langle n|\hat W|a\rangle}{E_a-E_n} 
\nonumber \\
               &+& \sum_n \frac{\langle b|\hat W|n\rangle\langle n|\hat D|a\rangle}{E_b-E_n},
\label{sos}
\end{eqnarray}
where $\hat D$ is electric-dipole operator and $\hat W$ is weak interaction operator.
In the RPA approximation, $\hat D= \hat H_{E1} + \delta \hat V_{E1}$ and
$\hat W= \hat H_{W}  + \delta \hat V_{W}$. Equation (\ref{sos}) is an
accurate expression for the E1 PNC amplitude if all states
$| a \rangle,| b \rangle,| n \rangle$ are exact many-electron states
of the whole atom. However, for our discussion we will consider only
states which differ from the ground state by excitation of the
single valence electron. This is reasonably good approximation
for alkali atoms and it is sufficient for the discussion.
In this case matrix elements in (\ref{sos}) are reduced to single-electron
matrix elements and we will refer to this expression as 
sum over single-electron states.
 
As one can see from Table~\ref{PNC}, Breit contribution to the correlation 
correction to the $s-s$ PNC amplitudes is small. This can be explained the 
following way. Breit and weak interactions are both short-range operators.
They have significant value in the vicinity of nucleus where all 
single-electron wave functions are similar and differ by normalization only.
Indeed, Dirac equation for all low states is the same on short distances
since energy can be neglected compared to the nuclear potential. On the
other hand, higher states with large energies do not contribute to the
sum (\ref{sos}) due to large energy denominator.
Therefore, relative value of Breit correction to all weak matrix
elements in (\ref{sos}) is the same and can be presented as a common
factor outside of the summation.
The effect of correlations on short distances can also be reduced to
change of normalization of the wave function. Therefore, relative Breit 
correction should remain the same.

This way of argument was used in calculating radiative corrections to
the PNC amplitude in cesium~\cite{Flambaum,Sushkov,Shabaev,Ginges2}.
Moreover, it was assumed in \cite{Sushkov} that to find radiative
correction it is sufficient to
consider only corrections to the weak matrix elements.
Radiative corrections to E1 transition amplitudes and energies
were not considered because it was believed that corresponding
contritions to the PNC amplitude are small and can be neglected.
It was demonstrated in \cite{Ginges3,Ginges2} that radiative
corrections to the PNC amplitude in cesium due to
change of energy denominators and change of E1 transition amplitudes
(see, eq. (\ref{sos})) are not so small when taken separately.
But they cancel each other almost exactly when taken together.

It is interesting to note that very similar situation takes 
place for the Breit correction. As it was first demonstrated
by Derevianko~\cite{Derevianko,Derevianko2} and then
confirmed by Dzuba {\em et al}~\cite{Harabati}, the effects of
Breit interaction on the PNC amplitude in cesium due to change
of energy denominators and change of E1 transition amplitudes
are not small, numerically very close, and have opposite signs.
Corresponding numbers are -0.4\%~\cite{Derevianko,Harabati}
for the effect on E1 amplitudes and 0.3\%~\cite{Derevianko}
and 0.4\%~\cite{Harabati} for the effect on energies.
There is almost exact numerical cancellation between two effects.
Similar situation takes place for francium.

However, as it is clear from the analysis of the $s-d$ PNC
amplitudes, numerical cancellation of the effect of Breit 
interaction on energies and E1 transition amplitudes is 
a fortunate feature of the $s-s$ PNC amplitudes rather than
a general situation. The same might be true for the radiative
corrections to the PNC amplitudes.

\begin{table}
\caption{Relative Breit correction to weak matrix
elements of Cs, Ba$^+$, Tl, Fr, and Ra$^+$.}
\label{weakme}
\begin{ruledtabular}
\begin{tabular}{lll}
Atom/Ion & $s-p$ & $p-d$ \\
\hline
Cs      &  -0.58\% &  -0.03\% \\
Ba$^+$  &  -0.65\% &  -0.29\% \\
Tl      &  -1.17\% &  -0.81\% \\
Fr      &  -0.98\% &  +1.60\% \\
Ra$^+$  &  -1.03\% &  +0.85\% \\
\end{tabular}
\end{ruledtabular}
\end{table}
  
In Table~\ref{weakme} we present the relative values of the effect
of Breit interaction on the $s-p$ and $p-d$ weak matrix elements 
between lowest valence states of five considered atoms. 
These states give dominant contribution to the PNC amplitudes.
Matrix elements are calculated in Brueckner approximation:
\[ W_{ab} = \langle \psi_b^{Br} | \hat H_W + \delta \hat V_W 
| \psi_a^{Br} \rangle. \]
Note, however, that relative value of the Breit correction to the $s-p$ 
weak matrix elements does not depend on the approximation. It is also
the same for states with different principal quantum number $n$ as far 
as the energy of this state is small compared to $-Ze^2/R_N$, where
$R_N$ is nuclear radius (see discussion above for more details).
Breit correction to the PNC amplitude
is approximately equal to the Breit correction to the weak
matrix element only for the $s-s$ PNC transitions of Cs and Fr. 
There is no obvious connection between the two 
for all other PNC amplitudes. This indicates a significant role
of the Breit correction to the E1 amplitudes and the energies
for the $p-d$ and $p_{1/2} - p_{3/2}$ PNC amplitudes (see eq.~(\ref{sos})).

Another distinctive feature of the $s-d$ PNC amplitudes which
can be easily seen from Table~\ref{PNC} is the significant and
non-trivial role of the correlations. One would expect that
correlations either play no role as for the $s-s$ PNC
amplitudes or enhance the effect of Breit interaction. 
Indeed, correlations increase electron density on short
distances which would lead to expectations that the
role of the Breit interaction should also increase. 
In fact, the opposite happens. As it is evident from Table~\ref{PNC} 
correlations decrease the Breit correction to the $s-d$ PNC amplitude.

To understand this behavior let us start our analysis 
from single-electron matrix elements of weak interaction. 
Arguments in favor of the small role
of correlations presented above for the $s-s$ PNC amplitudes
were based on the fact that both Breit and weak interactions
are localized on short distances. While it is true for the 
``bare'' operator of the weak interaction, it is not exactly
true when many-body effects are also included. The effective
operator of the weak interaction can be presented in the form
\begin{equation}
  \hat H^{\rm eff}_W = \hat H_W + \delta \hat V_W,
\label{heff}
\end{equation}
where first term is proportional to nuclear density and is
zero everywhere outside the nucleus. Second term is due to
change in self-consistent Hartree-Fock potential because
of the effect of weak interaction. It describes the so-called
core polarization effect and its radius is roughly equal to
the radius of atomic core. For the $s-p$ matrix elements
core polarization is just small correction to the leading
first term. Therefore all considerations based on assumption
that weak interaction is localized on short distances are still
approximately valid.

For the $p-d$ weak matrix elements first term in (\ref{heff})
gives zero for the point-like nucleus and all value for the matrix elements
is given by the second, core polarization term. Even for finite size 
nucleus, contribution of the first term in (\ref{heff}) is several
orders of magnitude smaller than of the second term. This means that
$p-d$ weak matrix elements don't come from short distances and all
arguments based on that assumption are not valid for them.

The $E1_{PNC}$ amplitude can be presented as a sum of three terms
(see eq.~(\ref{Ecor})). Corresponding numbers for the $s-d$ PNC
amplitudes in Cs, Ba$^+$, Fr and Ra$^+$ as well as for the 
$6p_{1/2} - 6p_{3/2}$ PNC amplitude in Tl are presented in Table~\ref{breit}.
The numbers in table are organized in such a way that the term which 
is proportional to the $s-p$ weak matrix element goes first, second
term is proportional to the $p-d$ matrix element and the last term
has the mix of both. Note that this last term is not included to the
sum over single-electron states expression~(\ref{sos}). It is small
for alkali atoms.
However, it is not negligible and we present it here to demonstrate
that its inclusion is important for accurate calculations.

\begin{table*}
\caption{Effect of Breit interaction on different contributions to the $s-d$ 
and $p_{1/2}- p_{3/2}$ $E1_{PNC}$ amplitudes in Brueckner approximation 
($10^{-11} iea_B Q_W/N$).}
\label{breit}
\begin{ruledtabular}
\begin{tabular}{lllcccc}
Atom & Transition & Breit &
$  \langle \psi_{a} | \hat H_{E1} + \delta \hat V_{E1}|\delta \psi_{b} \rangle $ &
$  \langle \psi_{a} | \hat H_{W}  + \delta \hat V_{W} | X_{b} \rangle  $ &
$  \langle \psi_{a} | \delta \hat V_{E1W}| \psi_{b} \rangle. \nonumber $ & Sum \\
or ion & \multicolumn{1}{c}{$a - b$} & (yes/no) & $\sim s-p$ weak m.e. & $\sim p-d$ weak m.e. & & \\
\hline
Cs & $5d_{3/2} - 6s_{1/2}$ & no  &  3.0045 & 0.8314 & -0.1081 &  3.7277 \\
   &                       & yes &  2.9846 & 0.8400 & -0.1081 &  3.7165 \\ 
   &                & $\Delta$   & -0.0199 & 0.0086 &  0.0000 & -0.0112 \\
   &             & $\Delta$ (\%) &  -0.66  & 1.0    &  0.0    & -0.30   \\
   &             &               &         &        &         &         \\
Ba$^+$ & $5d_{3/2} - 6s_{1/2}$ & no    &  2.5776 & -0.2486 & -0.1465 &  2.1826 \\
       &                       & yes   &  2.5583 & -0.2469 & -0.1462 &  2.1653 \\ 
       &               & $\Delta$      & -0.0193 &  0.0017 &  0.0003 & -0.0173 \\
       &               & $\Delta$ (\%) &  -0.75  &  0.07   &  0.02   & -0.79   \\
       &               &               &         &         &         &        \\
Tl &  $6p_{3/2} - 6p_{1/2}$ & no  & 37.454 & -1.195 & -11.989 & 24.270 \\
   &                        & yes & 37.106 & -1.191 & -11.869 & 24.046 \\
   &                 & $\Delta$   & -0.348 &  0.004 &   0.120 & -0.224 \\
   &               & $\Delta$(\%) & -0.93  &  0.03  &   1.0   & -0.92  \\
   &               &              &        &        &         &        \\
Fr & $6d_{3/2} - 7s_{1/2}$ & no  & 51.598 &  6.466 & -1.627 &  56.437 \\
   &                       & yes & 51.074 &  6.643 & -1.628 &  56.090 \\
   &                & $\Delta$   & -0.524 &  0.177 &  0.001 &  -0.347 \\
   &             & $\Delta$ (\%) & -1.00  &  2.74  &  0.0   &  -0.61  \\
   &             &               &        &        &        &         \\
Ra$^+$ & $6d_{3/2} - 7s_{1/2}$ & no  & 46.767 & -1.964 & -2.294 & 42.511 \\
       &                       & yes & 46.230 & -1.973 & -2.287 & 41.971 \\
       &                & $\Delta$   & -0.537 &  0.009 &  0.007 & -0.540 \\
       &             & $\Delta$ (\%) & -1.15  &  0.46  &  0.31   & -1.27 \\
\end{tabular}
\end{ruledtabular}
\end{table*}

The numbers in the table indicate that, as one would expect, term proportional
to the $s-p$ matrix element strongly dominates in the $E1_{PNC}$ amplitude.
However, the term proportional to the $p-d$ weak matrix element is not small
at all and brings significant complications to the behavior of the $E1_{PNC}$
amplitudes. The case of particular interest is comparison of the $s-d$ PNC
amplitudes of Cs and Ba$^+$. As one can see from Table~\ref{PNC} the role
of correlations in the Breit correction to these $E1_{PNC}$ amplitudes is
very much different in cases of Cs and Ba$^+$. Correlations almost do not
change Breit correction to the $s-d$ PNC amplitude in Ba$^+$.
On the other hand, their effect on the $s-d$ PNC amplitude in Cs is
significant. Moreover, it goes contrary to naive expectations as it was
discussed above. The Breit correction is smaller when correlations are
included while the effect of correlations on the density of external
electron on short distances suggests that it should be rather larger.
Another interesting thing is the difference between the results 
for Cs obtained in the $\hat \Sigma^{(2)}$ approximation 
(formula (\ref{final})) and Brueckner approximation (formula (\ref{Ecor})).
These two expressions differ by higher orders in $\hat \Sigma$. Therefore
the difference in results indicates important role of high-order $\hat \Sigma$
terms. This is an interesting fact given that the term linear in $\hat \Sigma$
strongly dominates in the correlation correction to the PNC amplitude.
Situation for the correlated Breit correction is opposite. The term
linear in $\hat \Sigma$ is small and higher-order terms dominate.

\begin{table*}
\caption{Effect of Breit interaction on
different contributions to the $5d_{3/2} - 6s$ PNC
amplitude in Cs and Ba$^+$ in TDHF approximation ($10^{-11} iea_B Q_W/N$).}
\label{s-d}
\begin{ruledtabular}
\begin{tabular}{llcccc}
Atom & Breit &
$  \langle \psi_{5d_{3/2}} | \hat H_{E1} + \delta \hat V_{E1}|\delta \psi_{6s} \rangle $ &
$  \langle \psi_{5d_{3/2}} | \hat H_{W}  + \delta \hat V_{W} | X_{6s} \rangle  $ &
$  \langle \psi_{5d_{3/2}} | \delta \hat V_{E1W}| \psi_{6s} \rangle. \nonumber $ & Sum \\
or ion & (yes/no) & $\sim s-p$ weak m.e. & $\sim p-d$ weak m.e. & & \\
\hline
Cs & no            &  2.9009 & 0.2806 & -0.0637 &  3.1177 \\
   & yes           &  2.8836 & 0.2814 & -0.0638 &  3.1012 \\
   & $\Delta$      & -0.0173 & 0.0008 &  0.0001 & -0.0165 \\
   & $\Delta$ (\%) &  -0.60  & 0.29   &  0.0    & -0.53   \\
   &               &         &        &         &          \\
Ba$^+$ & no            &  2.6079 & -0.3155 & -0.1245 &  2.1678 \\
       & yes           &  2.5886 & -0.3131 & -0.1242 &  2.1513 \\
       & $\Delta$      & -0.0193 &  0.0024 &  0.0003 & -0.0165 \\
       & $\Delta$ (\%) & -0.74   &  0.76   &  0.2    & -0.76   \\
\end{tabular}
\end{ruledtabular}
\end{table*}

For a more detailed analysis of the $s-d$ PNC amplitudes in Cs and 
Ba$^+$ we present another set of data in Table~\ref{s-d}. This data
is similar to those in Table~\ref{breit} but in TDHF approximation.
Data presented in Table~\ref{breit} is in Brueckner approximation 
which means that correlations are included to all orders in $\hat \Sigma$.
In contrast, data from Table~\ref{s-d} contains no correlations.
Comparison of the data in Table~\ref{breit} and Table~\ref{s-d} shows
that correlation correction to terms proportional to the $s-p$ weak 
matrix elements is small. This is similar to the $s-s$ PNC amplitude
and the same arguments are valid to explain this. On the other hand,
correlations significantly increase (by an order of magnitude) the
contribution of the terms proportional to the $p-d$ weak matrix
elements. Correspondingly, Breit correction also increases.
This is also well expected fact given that correlations increase
density of external electron in the core. In summary, different
behavior of the $s-p$ and $p-d$ weak matrix elements with respect
to correlations is due to different distances from which these
matrix elements get their values. The $s-p$ matrix elements are well
localized at atomic nucleus, while $p-d$ matrix elements localized
on much larger distances inside atomic core. 

\begin{table*}
\caption{Breit correction to the $6s - 5d_{3/2}$ PNC amplitude
in Cs in Brueckner approximation as a function of rescaling
parameter $S$ ($\hat \Sigma \rightarrow S \hat \Sigma$ in 
(\ref{Brab}) and (\ref{DBrab})) ($10^{-11} iea_B Q_W/N$).}
\label{SS}
\begin{ruledtabular}
\begin{tabular}{llcccc}
  $S$ &  No Breit & with Breit & $\Delta$ & $\Delta(S)-\Delta(0)$ &
  $\Delta(S=1)$ \\
\hline
 0.0 & 3.117696  &  3.101215  & -0.016481 &    0       &           \\
 0.1 & 3.160052  &  3.143462  & -0.016590 & -0.000109  & -0.017571 \\
 0.2 & 3.203363  &  3.186710  & -0.016653 & -0.000172  & -0.017341 \\
 0.4 & 3.295216  &  3.278633  & -0.016583 & -0.000102  & -0.016736 \\
 0.6 & 3.400763  &  3.384702  & -0.016061 &  0.000420  & -0.015781 \\
 0.8 & 3.534769  &  3.520128  & -0.014641 &  0.001840  & -0.014181 \\
 1.0 & 3.727607  &  3.716356  & -0.011251 &  0.005230  & -0.011251 \\
\end{tabular}
\end{ruledtabular}
\end{table*}

Since correlations have different effect on Breit correction
to the $6s_{1/2} - 5d_{3/2}$ PNC amplitude in Cs depending on
whether they are included in linear in $\hat \Sigma$ approximation 
(formula (\ref{final})) of in all-order Bruekner approximation 
(formula (\ref{Ecor})), it is important to check
that all difference is due to higher order in $\hat \Sigma$ terms.
To do this we extract linear in $\hat \Sigma$ contribution
from Brueckner approximation and compare it with the result obtained
using formula (\ref{final}). This is done with the use of rescaling
procedure similar to those discussed in the end of previous section for 
the Breit operator. However, this time we rescale correlation potential
$\hat \Sigma$. Namely, we replace $\hat \Sigma$ in eqs. (\ref{Brab})
and (\ref{DBrab}) by $S \hat \Sigma$, where $S$ is rescaling
parameter. Calculating Breit correction with $S \ll 1$ and extrapolating
it to $S=1$ leaves only linear in $\hat \Sigma$ contributions.
The results are presented in Table~\ref{SS}. 
Extrapolation to $S=1$ is done using the formula
\[ \Delta(S=1) = \Delta(0) + \frac{\Delta(S) - \Delta(0)}{S}. \]
Corresponding numbers are presented in the last column of Table~\ref{SS}.
One can see that
the value of Breit correction calculated at $S=0.1$ and extrapolated to
$S=1$ (-0.017571) practically coincide with the value -0.0176 obtained 
using formula (\ref{final}) (see Table~\ref{PNC}). For larger values
of $S$ non-linear in $\hat \Sigma$ terms become important and rapidly
change the value of Breit correction.

There are few more important conclusions which can be drawn from the
fact that Breit correction to a PNC amplitude is only sensitive to
change of the electron density. This change of density is
proportional to the value of the correlation potential $\hat \Sigma$.
Therefore, it is important to check how inclusion of some extra
terms into $\hat \Sigma$ affect the value of the Breit 
correction to the PNC amplitudes. For example, inclusion of the
Breit interaction itself into calculation of $\hat \Sigma$
does not change its value very much. Corresponding change of removal
energies of Cs is less than 0.02\%, while change of the
$6s - 6p_{1/2}$ energy interval is 0.04\% only. Test 
calculations show that corresponding change to the PNC
amplitude is also small, it is 0.04\% for the $6s-5d{3/2}$ PNC 
amplitude and even smaller for the $6s - 7s$ PNC amplitude.
On the other hand, it is well known from a number of calculations
(see, e.g.~\cite{all-order}) that inclusion into $\hat \Sigma$
of certain higher-order Coulomb terms leads to significant change
in the value of $\hat \Sigma$ and electron density on short
distances, brining theoretical values for energies, hyperfine 
structure, transition amplitudes, etc. to much better agreement 
with the experiment. Therefore, it would be inappropriate 
to assume {\em a priori} that higher-order Coulomb terms
do not affect Breit correction. This is to be checked numerically.
We have performed such calculations and the results show that
the effect of the higher-order Coulomb terms in $\hat \Sigma$
is indeed small. Its value is close to the effect of inclusion
of Breit interaction into $\hat \Sigma$. Moreover, the two
effects tend to cancel each other. One can say therefore
that inclusion of Breit and higher-order Coulomb terms 
into $\hat \Sigma$ are equally important or unimportant
depending of the needed accuracy.

Let us now come back to comparison of the $s-d$ PNC amplitude for
Cs and Ba$^+$ for few more important observations.
If terms proportional to $s-p$ and $p-d$ weak matrix elements are
compared separately, their behavior is very similar for both
atoms. This is a natural consequence of the similar electron
structure. However, the total $s-d$ PNC amplitude is more
sensitive to small difference in the electron structure.
In Ba$^+$ ion $5d$ states are lower than the $6p$ states
while in Cs  $5d$ states are higher than the $6p$ states.
This leads to larger correlation correction to the $p-d$
part of the PNC amplitude in Ba$^+$ compared to those
of Cs. This also leads to the opposite sign of these terms
in Ba$^+$ and Cs. As a result, the effect of correlations on
final PNC amplitude is very different. In general, the
effect of correlations on $s-d$ and $p_{1/2} - p_{3/2}$
amplitudes is very hard to predict due to different behavior 
of terms proportional to $s-p$ and $p-d$ weak matrix elements.

Having accurate results for Breit correction to the PNC amplitudes
for many atoms we can now check how they rescale with the nuclear
charge $Z$. This would allow us to obtain Breit correction for
other atoms without doing sophisticated calculations.
It is easy to see that Breit correction to similar amplitudes
in atoms with similar electron structure is just proportional to $Z$.
For example, $\Delta E1_{PNC}({\rm Fr}) = (87/55)\Delta E1_{PNC}({\rm Cs})$.
It is interesting to note that rescaling from the $6s-7s$
amplitude in Cs to the $6p_{1/2}-6p_{3/2}$ amplitude in Tl gives
result which is very close to the correct answer:
$\Delta E1_{PNC}({\rm Tl}) = (81/55)\Delta E1_{PNC}({\rm Cs})=-0.91\%$
while accurate calculations give $-0.92\%$. However, this is just
fortunate coincidence. Separate analysis of terms proportional to
the $s-p$ and $p-d$ weak matrix elements show that no reliable rescaling
between Cs and Tl is possible. The same is true for any other
pair of the PNC amplitudes which correspond to different
electron structures.

\begin{table}
\caption{Breit corrections to the PNC amplitudes
in Cs, Ba$^+$, Tl, Fr and Ra$^+$ ($10^{-11} iea_B Q_W/N$),
comparison with other calculations.}
\label{comp}
\begin{ruledtabular}
\begin{tabular}{lllll}
Atom & Transition & $\Delta_{\rm Breit}$  & Source \\
\hline
Cs & $6s - 7s$ & -0.0056         & this work \\
           &           & -0.0054         & Derevianko~\cite{Derevianko2} \\
           &           & -0.0055         & Dzuba {\em et al} \cite{Harabati} \\
           &           & -0.004          & Kozlov {\em et al} \cite{Kozlov} \\
           &           & -0.0045         & Shabaev {\em et al} \cite{Shabaev1} \\
     & $6s - 5d_{3/2}$ & -0.0112           & this work \\
Ba$^+$ & $6s - 5d_{3/2}$ & -0.0172 & this work \\
Tl & $6p_{1/2}-6p_{3/2}$ & -0.224         & this work \\
Fr & $7s - 8s$ & -0.15           & this work \\
           &           & -0.18           & Derevianko~\cite{Derevianko3} \\
           &           & -0.15           & Safronova  \\
           &           &                 & and Johnson~\cite{Safronova} \\
           &           & -0.14           & Shabaev {\em et al} \cite{Shabaev} \\
Ra$^+$ & $7s - 6d_{3/2}$ & -0.541         & this work \\
\end{tabular}
\end{ruledtabular}
\end{table}

We present our final results in Table~\ref{comp} together
with the results of other authors. The results for the $6s-7s$ 
amplitude in Cs and $7s-8s$ amplitude in Fr are in good agreement 
with other calculations. However, we believe that our results are
the most accurate due to non-perturbative treatment of the Breit
interaction which is complete up to second order of the MBPT.
The results for other atoms as well as for the $s-d$ PNC 
amplitudes for Cs and Fr are presented for the first time.

We now can use the result of this work for the Breit interaction contribution
and our previous calculations of other contributions to find the PNC
amplitude in $^{133}$Cs, extract value of the weak charge and compare it with 
the standard model. Our  many-body calculations of the PNC amplitude
produced by the electron-nucleus weak interaction gave the following
result \cite{Dzuba02} (without the Breit contribution and QED radiative
corrections)\\
$E_{PNC}= -0.9060 (1 \pm 0.5\%)\times 10^{-11} iea_{B}(-Q_{W}/N)$ .\\
We take the Breit contribution -0.0056 from this work and the QED radiative
corrections -0.0029 (-0.32 \% of the PNC amplitude) from Ref. \cite{Ginges2}.
This gives 
\begin{equation}\label{EPNC}
E_{PNC}= -0.8975 (1 \pm 0.5\%)\times 10^{-11} iea_{B}(-Q_{W}/N) \ .
\end{equation}
In the experiment \cite{Wood}  the ratio
$E_{PNC}/\beta$, where $\beta$ is the vector transition polarizability, 
actually has been measured. To extract $Q_w$ we used
  $\beta=26.99(5) a_B^3$ obtained in  \cite{Dzuba02}
as the statistical average   of  the two most  accurate values
of $\beta$.
From the measurements of the PNC amplitude \cite{Wood} we obtain
\begin{equation}\label{Qw}
Q_W=-72.69(29)_{\rm exp}(36)_{\rm theor} \ .
\end{equation}
The difference with the standard model value $Q_W^{\rm SM}=-73.19(3)$
 \cite{Data} is
\begin{equation}\label{DeltaQ}
Q_W-Q_W^{\rm SM}=0.50(46) \ ,
\end{equation}
adding the errors in quadrature. Thus, the difference between the central
points is one standard deviation.

For $^{205}$Tl we use PNC amplitude from Ref. \cite{CPM}, \\
$E_{PNC}= -(27.0  \pm 0.8 )\times 10^{-11} iea_{B}(-Q_{W}/N)$,\\
the QED radiative correction -0.57 \% (see detailes in \cite{Ginges}), 
the neutron skin correction -0.3 \% \cite{Kozlov1}
and the Breit correction -0.92 \% from this work. The result 
is 
\begin{equation}\label{EPNCTl}
E_{PNC}= -(26.5 \pm 0.8) \times 10^{-11} iea_{B}(-Q_{W}/N) \ .
\end{equation}
Using M1 amplitude 1.693 a.u. from \cite{Kozlov1}
and measurement \cite{Seattle}
 $Im(E_{PNC}/M1)=-(14.68 \pm 0.17) \times 10^{-8}$ a.u.
we  obtain for $^{205}$Tl
\begin{equation}\label{QwTl}
Q_W=-116.2(1.3)_{\rm exp}(3.5)_{\rm theor} \ .
\end{equation}
The standard model value is -116.81(4) \cite{Data}.
 The difference with the standard model for Tl is
\begin{equation}\label{DeltaQTl}
Q_W-Q_W^{\rm SM}= 0.6(3.7) \ ,
\end{equation}

\section{Conclusion}

We have developed a method of non-perturbative treatment of the Breit
interaction in accurate calculations for many-electron atoms.
The method has been applied to calculation of the Breit correction
to the PNC amplitudes of Cs, Ba$^+$, Tl, Fr and Ra$^+$.
The results for the $6s-7s$ 
amplitude in Cs and $7s-8s$ amplitude in Fr are in good agreement 
with other calculations. The results for other atoms and for the 
$s-d$ PNC  amplitudes for Cs and Fr are presented for the first time.
The following features of the Breit correction
to the PNC amplitudes have been revealed:
\begin{itemize}
\item The effect of correlations on the $s-s$ PNC amplitudes is small.
\item In contrast, correlations are important for those PNC
amplitudes which depend on the $p-d$ weak matrix elements between 
valence states. 
\item For accurate treatment of correlations inclusion of linear
in the correlation potential $\hat \Sigma$ terms is not enough.
The main effect is due to the change in density of external electron
inside the core which is magnified in higher orders in $\hat \Sigma$.
It can be included by the use of Brueckner orbitals.
\item Inclusion of the Breit interaction as well as higher-orders 
Coulomb terms into correlation potential $\hat \Sigma$ have only small
effect on PNC amplitudes and can be neglected in most cases.
\item Numerical cancellation of the effects of Breit interaction 
on E1 transition amplitudes and energy denominators in the $s-s$ PNC 
amplitudes for Cs and Fr is rather fortunate. No such cancellation
takes place for other atoms and for the $s-d$ amplitudes of Cs and Fr.
The same is probably true for the radiative corrections to the PNC
amplitudes.
\item Rescaling of the Breit correction to the PNC amplitudes 
using ratio of the nuclear charges $Z$ can only be done to similar 
amplitudes in atoms with similar electron structure.
\end{itemize}

Combining Breit corrections to PNC amplitudes in cesium and thallium
with previous accurate calculations and measurements confirms good
agreement of the weak nuclear charges of these atoms with the prediction
of the standard model.


\end{document}